\documentclass[aps,prl,preprint,superscriptaddress,showpacs,preprintnumbers]{revtex4}
\usepackage{graphicx}
\usepackage{dcolumn}
\begin{document}
\title{From chiral vibration to static chirality in $^{135}$Nd}
\author{S.~Mukhopadhyay}
\affiliation{Physics Department, University of Notre Dame, Notre Dame, IN
46556, USA}
\affiliation{UGC-DAE Consortium for Scientific Research, Kolkata Centre,
Kolkata 700098, India}
\author{D.~Almehed}
\author{U.~Garg}
\author{S.~Frauendorf}
\author{T.~Li}
\author{P.V.~Madhusudhana~Rao}
\affiliation{Physics Department, University of Notre Dame, Notre Dame, IN
46556, USA}
\author{X.~Wang}
\affiliation{Physics Department, University of Notre Dame, Notre Dame, IN
46556, USA}
\affiliation{Physics Division, Argonne National Laboratory, Argonne, IL
60439, USA}
\author{S.S.~Ghugre}
\affiliation{UGC-DAE Consortium for Scientific Research, Kolkata Centre,
Kolkata 700098, India}
\author{M.P.~Carpenter}
\author{S.~Gros}
\affiliation{Physics Division, Argonne National Laboratory, Argonne, IL
60439, USA}
\author{A.~Hecht}
\affiliation{Physics Division, Argonne National Laboratory, Argonne, IL
60439, USA}
\affiliation{Department of Chemistry and Biochemistry, University of Maryland,
College Park, MD 20742, USA}
\author{R.V.F.~Janssens}
\affiliation{Physics Division, Argonne National Laboratory, Argonne, IL
60439, USA}
\author{F.G.~Kondev}
\affiliation{Nuclear Engineering Division, Argonne National Laboratory, Argonne,
IL 60439, USA}
\author{T.~Lauritsen}
\author{D.~Seweryniak}
\author{S.~Zhu}
\affiliation{Physics Division, Argonne National Laboratory, Argonne, IL
60439, USA}
\date{\today}

\begin{abstract}
Electromagnetic transition probabilities have been measured for
the intra- and inter-band transitions in the two sequences in the nucleus
$^{135}$Nd that were previously identified as a composite chiral pair of
rotational bands.
The chiral character of the bands is affirmed and it is shown that
their behavior is associated with a transition from a vibrational into a 
static chiral regime.

\end{abstract}


\maketitle
Chirality is a well-known phenomenon in chemistry and biology as a geometric
property of many molecules, in particular of complex biomolecules.
Particle physics is another domain of
chirality, where it describes a kinematical feature of massless
particles.
Both in chemistry and particle physics, space inversion
changes left-handed into right-handed systems.
Nuclei have been considered as
achiral, because their shapes are, generally, too simple. However, it was
pointed out some time ago that a triaxial nucleus
becomes chiral if it rotates about an axis that lies outside the three
planes
spanned by the principal axes of its triaxial ellipsoidal shape
\cite{Stefan1,Stefan2}. The short,
intermediate, and long axes form a screw with respect to the angular
momentum
vector. In contrast with molecules and massless particles, space inversion
leaves nuclear chirality unchanged.
The left-handed configuration is converted into the right-handed one by
the time reversal operation, which changes the sign of all linear and 
angular momenta. Since both chiral structures have the
same energy and are related by time reversal,
one expects to observe two identical  bands of the same parity. A
number of pairs of bands have been identified in nuclei in the 
A$\sim$130 and
A$\sim$100 regions of the nuclear chart and have been suggested as 
candidates for chiral partners
\cite{Petrache1, Starosta1, hecht, koike1, hart, zhu, vaman, joshi1, timar}.
A small
observed energy difference between the states of the same spin, $I$, in 
the chiral partners indicates rapid conversion between the left- and 
right-handed
configurations (chiral vibration). With decreasing energy splitting between
the partner bands, the left-right
mode changes from soft chiral vibration to tunneling between 
well-established chiral configurations (static chirality).

In addition to close level energies, the two chiral partners should de-excite
in a very similar way via electromagnetic radiation. Thus, in this 
context, a measurement of $E2$ transition probabilities is an essential 
probe of nuclear chirality.
Only two lifetime measurements have been published so far. For $^{128}$Cs
\cite{grod}, it was established that the electromagnetic transition
probabilities are similar; but, the
 excitation energies of the two partner bands never approach each other.
 In  $^{134}$Pr, the two bands come very close around spin 14; however,
  the intra-band $B(E2)$ values were found to differ about by a factor of two  
\cite{Tonev,Petrache2}.
The latter observation was interpreted as suggesting that the change in
orientation
of the angular momentum vector must be accompanied by a change in shape
\cite{Tonev},
or even that the results rendered the chiral interpretation itself doubtful
\cite{Petrache2}.

The study of magnetic rotation \cite{Stefan1} suggests that chirality is 
better developed if the non-planar
geometry is generated by three excited quasiparticles (instead of two, as in
case of $^{134}$Pr and $^{128}$Cs). 
Such a case was realized in the nucleus
$^{135}$Nd, where two partner bands consistent with chirality have been
observed \cite{zhu}. This expectation is now borne out by the 
measurements and
calculations presented below as the electromagnetic transition
rates in the chiral bands of $^{135}$Nd were found to be nearly identical.
For the first time, the splitting between the chiral partners is 
calculated in a microscopic way by
extending the tilted-axis-cranking model (TAC) by the random
phase approximation (RPA).
The observed transition probabilities and energies are in very good
agreement with the calculations, which indicate, furthermore, that a 
transition from chiral vibration to static chirality occurs with increasing
spin.

The experiment was carried out at the ATLAS accelerator facility at the 
Argonne National
Laboratory, and employed a 175-MeV $^{40}$Ar beam to populate high-spin
states of $^{135}$Nd with the $^{100}$Mo($^{40}$Ar, 5n) reaction. The
target was
a 1.14 mg/cm$^{2}$-thick, isotopically-enriched, $^{100}$Mo foil backed by a
17.9 mg/cm$^{2}$-thick layer of Pb in order to slow down and stop the
recoiling nuclei.
A total of about 2.5 $\times$ 10$^{9}$ five- and higher-fold coincidence events
were accumulated using the Gammasphere array \cite{Lee}.
The level scheme of $^{135}$Nd is already well-established \cite{piel,Beck}
and the chiral bands were presented in Ref. \cite{zhu}; in the remainder 
of this
paper, we refer to the partial level scheme presented therein.

For the analysis of these thick-target data with the Doppler-Shift 
Attenuation
Method (DSAM), the power of the BLUE database approach \cite{Cromaz} has 
been
employed to efficiently sort the data angle-by-angle. After building the
database
for each fold, and performing appropriate background subtraction
\cite{Starosta4},
double-gated coincidence spectra were obtained for each ring (angle) of
Gammasphere
for further analysis of the relevant Doppler shifts. In the cases where
the gating
transition was not fully stopped, the gate was made sufficiently wide to
ensure that
the full peak shape was included in the gate; this eliminated a possible
bias in the analysis that could be introduced by omitting part of a 
peak's lineshape, thereby favoring a specific time component of the
gating transition.

Lifetimes of states in both chiral bands were extracted using the
LINESHAPE~analysis codes of Wells and Johnson \cite{Wells}. In each case, a
total of 5000 Monte Carlo simulations of the velocity history of the
recoiling
nuclei traversing the target and the backing material were generated in time
steps
of 0.002 ps. Electronic stopping powers were calculated with the code 
SRIM~\cite{Ziegler}. More details about the fitting procedures can be
found in Refs. \cite{Chiara1,Chiara2}.

\begin{figure}[htp]
\includegraphics[angle=270,width=8.6cm]{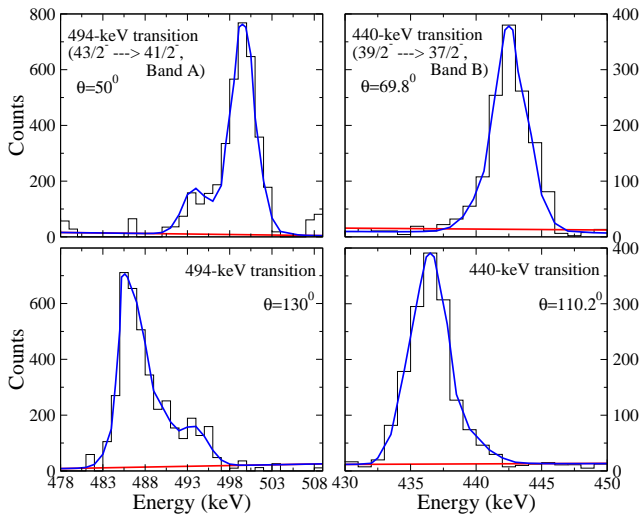}
\caption{\label{fig:fit}(Color online) Lineshape fits for representative
$\gamma$ transitions in the chiral bands of $^{135}$Nd. Left: the
494-keV transition in Band A. Right: the 440-keV transition in Band B.}
\end{figure}

The lifetimes of the levels with spins, $I$, from 29/2$^-$ to 43/2$^-$ 
in Band A, and 
from 31/2$^-$ to 39/2$^-$ in Band B, have been deduced in this work.
Representative
examples of DSAM fits are displayed in Fig.~\ref{fig:fit}, and the extracted
lifetimes are
listed in Table I. Uncertainties in the lifetimes were determined from
the behavior of the
$\chi$$^2$-fit in the vicinity of the minimum
\cite{Chiara1,Chiara2,Johnson}.

\begin{table}
\vskip -0.2cm
\caption{\label{tab:table1}Derived lifetimes and electromagnetic
transition probabilities in the chiral bands of $^{135}Nd$}
\begin{ruledtabular}
\begin{tabular}{cccc}
&\multicolumn{3}{c}{Band A}\\
\cline{2-4}
Spin I $[\hbar]$ & Lifetime[ps] & $B(M1)[\mu_n^2]$ & $B(E2)[e^2b^2]$\\ 
\hline
\hline
$\frac {29}{2}^{-}$ &  1.00(5)  &   3.2(2) &   0.32(2)\\
$\frac {31}{2}^{-}$ &  0.75(8)  &   2.5(3) &   0.32(3)\\
$\frac {33}{2}^{-}$ &  0.44(2)  &   2.2(2) &   0.32(3)\\
$\frac {35}{2}^{-}$ &  0.28(2)  &   2.4(3) &   0.32(3)\\
$\frac {37}{2}^{-}$ &  0.23(2)  &   1.7(3) &   0.32(4)\\
$\frac {39}{2}^{-}$ &  0.22(2)  &   2.1(3) &   0.13(3)\\
$\frac {41}{2}^{-}$ &  0.18(2)  &   2.1(3) &   0.19(3)\\
$\frac {43}{2}^{-}$ &  0.16(2)  &   2.0(3) &   0.21(4)\\
\hline
&\multicolumn{3}{c}{Band B}\\
\cline{2-4}
$\frac {31}{2}^{-}$ &  1.46(2)  &  2.7(3) & 0.28(3)\\
$\frac {33}{2}^{-}$ &  0.87(6)  &  2.1(2) & 0.28(3)\\
$\frac {35}{2}^{-}$ &  0.64(5)  &  2.2(2) & 0.28(4)\\
$\frac {37}{2}^{-}$ &  0.48(3)  &  1.7(2) & 0.29(4)\\
$\frac {39}{2}^{-}$ &  0.24(4)  &  1.9(3) & 0.11(3)\\
\end{tabular}
\end{ruledtabular}
\vskip -0.4cm
\end{table}

\begin{figure}
\includegraphics[angle=270,width=8.6cm]{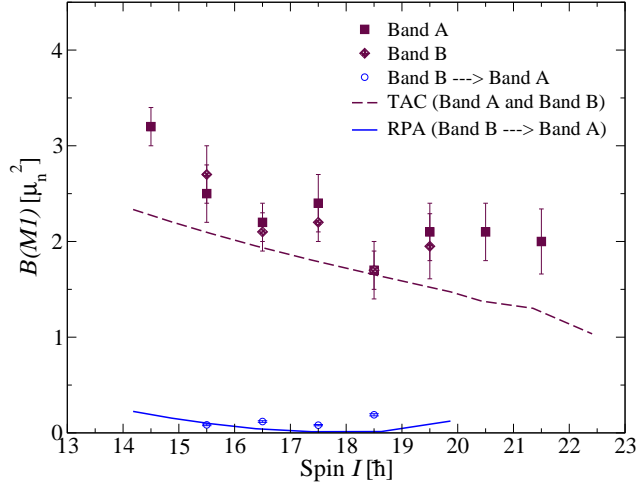}
\caption{\label{fig:bm1}(Color online) Evolution of the $B(M1)$ transition
rates with
spin for the two chiral partner bands in $^{135}$Nd. A comparison with
the calculations described in the text is shown as well.}
\end{figure}

\begin{figure}[htp]
\includegraphics[angle=270,width=8.6cm]{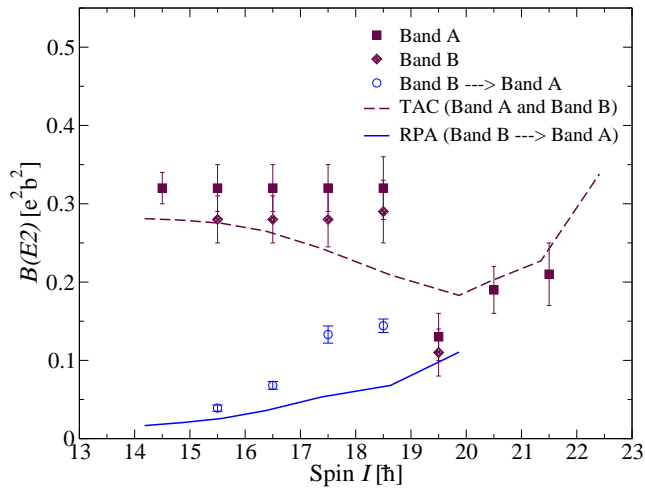}
\caption{\label{fig:be2}(Color online) Variation of the $B(E2)$ transition
rates with
spin for the two chiral partner bands in $^{135}$Nd, together with the
results of the calculations described in the text.}
\end{figure}

Reduced transition probabilities, $B(M1)$ and $B(E2)$, were
derived
from the resulting lifetimes and are presented in Figs.~\ref{fig:bm1}
and~\ref{fig:be2}. Within the experimental uncertainties, the $B(E2)$
and $B(M1)$ values
for the intra-band transitions of the two bands are essentially the
same, pointing to their
identical nature. In addition, the $B(M1)$ values of
the intra-band
transitions exhibit a characteristic staggering with increasing spin. A 
similar
staggering, but opposite in phase, is observed in the $B(M1)$
values of the inter-band transitions.
All these results
are fully compatible with a pure chiral interpretation of these bands 
~\cite{Stefan2,koike3}.

We have performed new calculations for the twin bands in $^{135}$Nd
in the framework of TAC complemented by RPA.
As proposed in Ref.~\cite{zhu}, these bands 
are built on the $\{\pi h^2_{11/2}, \nu h_{11/2}\}$
configuration. TAC is a microscopic mean field method that has been shown 
to describe the energy and the inband transition rates of
the lower of the two chiral partner bands very well (see, {\em e.g.},
Refs. \cite{Dimitrov,hecht,timar}).
However, to describe the energy
splitting between the two partner bands, one needs to go beyond the mean field
approach. This has previously been done only in either the
2-particle-plus-rotor~\cite{Stefan2,bark} or the core-quasiparticle-coupling
model \cite{Starosta3}.
But, these models cannot be applied to the 3-quasiparticle
configuration of $^{135}$Nd. Moreover, both these approaches assume that
the moments of inertia are of the irrotational-flow type, which is not 
consistent with microscopic cranking calculations~\cite{Tonev}.
RPA, as a microscopic approach, does not have these restrictions since it
describes the excitations of the equilibrium mean field as harmonic
vibrations.


We use the self-consistent TAC with the QQ-force and a
constant pair gap (PQTAC)~\cite{Stefan3} in two major N-shells (4 and 5):

\begin{equation}
H'=h_{0}+\sum_{\mu,N} \frac{\kappa_N}{2} Q^{(N)}_\mu Q^{(N)}_\mu (-)^\mu 
-\Delta \left( P^+ + P \right) - \vec{\omega} \cdot \vec{J}
\end{equation}
where $h_{0}$ is the spherical Woods-Saxon energy~\cite{Dimitrov} and
$Q^{(N)}_\mu$ are the 5 quadrupole operators
for each N-shell with an N-dependent force strength~\cite{Baranger}:
\begin{equation}
\kappa_N = \kappa_0 \frac{N_L-B}{N-B} \left(\frac{2 Z(N)}{A}\right)^{1/3}.
\end{equation}
Here, $N_L=4$ and the parameters $\kappa_0$ and $B(=0.5)$ are chosen to 
reproduce the Strutinsky
potential energy surface. The rotational frequency in the cranking term,
$\vec{\omega} \cdot \vec{J}$, is defined as
$\omega_1=\omega\sin\vartheta\cos\varphi$,
$\omega_2=\omega\sin\vartheta\sin\varphi$  and
$\omega_3=\omega\cos\vartheta$, where $\vartheta$ and $\varphi$ are the two
tilt angles defined in Ref.~\cite{Dimitrov}.
We use a proton (neutron) pairing gap, $\Delta_{n,p}$, that is 50\%
(80\%) of
the odd-even mass difference. Since the proton configuration 
corresponds to a particle-hole excitation of the even-even system, we have used 
a smaller pairing gap than that of an odd-neutron system.

The RPA calculates the harmonic excitations around the mean field
minimum. This chiral vibrational regime applies
 well below the transition point to chiral rotation,
where the RPA solution goes to zero energy and the tilted mean field
gets a nonzero $\varphi$ value.
We solved the RPA equations:
\begin{equation}
\left[H_{RPA}, O^+_\lambda \right]= E_{RPA} O^+_\lambda
\end{equation}
where $H_{RPA}$ is the Hamiltonian in RPA order~\cite{ring}
derived from the corresponding self-consistent mean field wave function,
and $O^+_\lambda$ are the RPA eigenmode operators. 
$H_{RPA}$ describes the orientation-, shape-, 
and two quasi-particle degrees of freedom in the harmonic approximation.
The RPA phonon energy
gives the energy splitting between the zero-phonon lower band and the
one-phonon excited band, and the RPA wave functions
give the inter-band transition rates.
 Since RPA describes harmonic vibrations around the equilibrium, 
it does not contribute to the expectation values of the operators, {\em i.e.},
the
intra-band transition rates remain the same as those given by the TAC model. 
Details of the RPA calculations are provided in Ref. \cite{daniel}.

\begin{figure}[htp]
\includegraphics[angle=270,width=8.6cm]{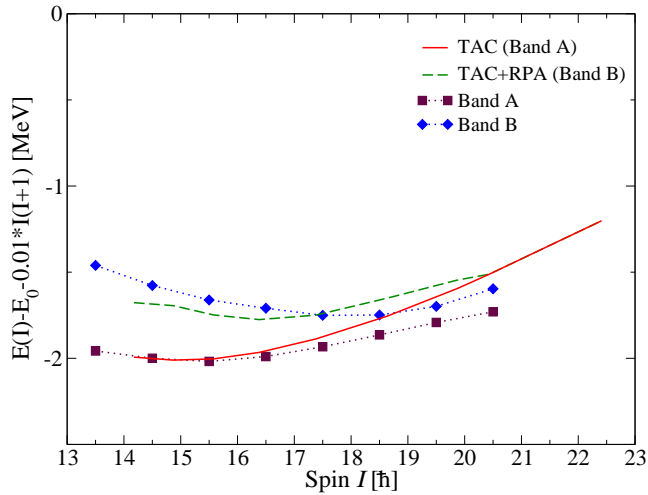}
\caption{\label{fig:rotor-ref}(Color online) Comparison, as a function of spin,
of the measured and calculated energies, relative to the band head, $E_{0}$,
of the chiral bands, with a rotor reference subtracted.}
\end{figure}

The TAC results reproduce the rotational energies of the bands rather well, 
as can be seen in Fig.~\ref{fig:rotor-ref}, and the one-phonon energy
matches the energy splitting between the two bands.

The  angular momentum where the RPA energy goes to zero and  static 
chirality emerges
in the TAC calculations is one unit  higher than the point
at which the two bands come closest in the data. This coming together of 
the two bands has been interpreted as the onset of chiral rotation 
\cite{zhu}. By analyzing the RPA wave functions, we can 
determine that the lowest phonon is dominated by orientation fluctuations 
in the orientation of the shape relative to the angular momentum vector, with a small
admixture of $\gamma$-vibration. Details of this analysis also  
are provided in Ref.~\cite{daniel}.

The calculated intra- and inter-band transition
probabilities are also displayed in
Figs.~\ref{fig:bm1} and \ref{fig:be2}, respectively. There is a good 
agreement with the data.
Below the critical frequency,  the TAC calculations give $\varphi=0$ 
and a slow increase of $\vartheta$, resulting in a slowly decreasing 
intra-band $B(E2)$. After the 
critical frequency, the rapid increase of $\varphi$ causes the steep 
rise in the $B(E2)$.
The calculated
inter-band $B(E2)$'s are somewhat smaller than the observed values but 
exhibit
the correct functional form. In accordance with the experiment,
the calculated inter-band $B(M1)$ rates are much smaller
than the intra-band $B(M1)$'s; so the observed different behavior with 
spin is
not of much consequence. Thus, the calculations reproduce the data rather
well, even close to the point of transition to static chirality, where the
TAC+RPA approach breaks down.

The nucleus $^{135}$Nd (3 quasiparticles) differs from  
$^{134}$Pr (2 quasiparticles) by the additional $h_{11/2}$
quasiproton.  The $B(E2)$ values within the two bands
are about the same, as long as the chiral vibration
remains near-harmonic. As will be shown in a forthcoming paper \cite{daniel},
they may differ substantially near the instability point,
where strong anharmonicities make the average tilt-angle, $\vartheta$,
different in the two bands. 
The resulting difference in the $B(E2)$ values is most pronounced
near $\vartheta = 45^o$, where the $B(E2)$'s go to zero.
As compared to $^{134}$Pr, the longer proton angular 
momentum in $^{135}$Nd delays this instability, 
{\em i.e.}, the chiral vibration is near-harmonic 
for a longer spin range. The additional quasiproton also leads to an
increase in $\vartheta$ from $\sim50^o$
to $\sim70^o$. These two effects explain 
why the $B(E2)$ values are different in $^{134}$Pr, but almost the same in
$^{135}$Nd.

In summary, we have measured electromagnetic transition probabilities for
intra- and inter-band transitions in the two bands in 
$^{135}$Nd
that were previously identified as a composite chiral pair.
The intra-band transition probabilities in the two bands are nearly 
identical, establishing them as chiral partners.
A microscopic calculation based on the combination of TAC with RPA
reproduces all experimental observables quite well, substantiating the
theoretical interpretation: At the bottom of the bands, the angular 
momentum vector
oscillates perpendicular to the plane spanned by the long and short axes 
of the
triaxial nuclear shape
(chiral vibration). These oscillations slow down with increasing angular
momentum, resulting in a decreasing energy splitting between the bands and
an increase in the inter-band $B(E2)$ values. The vibration, then, 
becomes strongly anharmonic,
changing into tunneling between well established  left- and right-handed
configurations (chiral rotation),
the best realization of chirality in nuclei known so far.

One of the authors (SM) expresses his gratitude to Dr. C. J. Chiara 
for several helpful discussions on the LINESHAPE codes.
This work has been supported in part by the U.S. National
Science Foundation (grants No. PHY04-57120 and INT-0111536);
by the Department of Science and Technology, Government of
India (grant No. DST-NSF/RPO-017/98); by the U.S. Department of
Energy, Office of Nuclear Physics, under contracts No. DE-AC02-06CH11357 and
DE-FG02-95ER40934; and, by the University of Notre Dame and the ANL-UND Nuclear
Theory Initiative.
\bibliography{ndchiral}
\clearpage
\end{document}